\documentclass[10pt, conference, compsocconf]{IEEEtran}
\usepackage[nolist]{acronym}

\usepackage[utf8]{inputenc}

\ifCLASSINFOpdf
	\usepackage[pdftex]{graphicx}
\else
\fi
\usepackage[cmex10]{amsmath}
\usepackage{amssymb}
\usepackage{algorithmic}
\begin{document}

\onecolumn
\thispagestyle{empty}

\par\noindent\textbf{This paper is a preprint (IEEE ``accepted'' status).}

\bigskip\par\noindent
\textbf{IEEE copyright notice.}~\mbox{\copyright{}} 2011 IEEE. Personal use of this material is permitted. Permission from IEEE must be obtained for all other uses, in any current or future media, including reprinting/republishing this material for advertising or promotional purposes, creating new collective works, for resale or redistribution to servers or lists, or reuse of any copyrighted component of this work in other works.	
\bigskip\par\noindent
\textbf{DOI.}  10.1109/CCP.2011.22
\bigskip\par\noindent
http://doi.ieeecomputersociety.org/10.1109/CCP.2011.22
\twocolumn

\begin{acronym}
	\acro{PPM}{Prediction by Partial Matching}
	\acro{DMC}{Dynamic Markov Coding}
	\acro{CTW}{Context Tree Weighting}
	\acro{PAQ}{``Pack''}
	\acro{AC}{Arithmetic Coding}
	\acro{CM}{Context Mixing}
	\acro{CG}{Conjugate Gradient}
	\acro{KT}{Krichevsky-Trofimov}
	\acro{iid}{independent identically distributed}
	\acro{BWT}{Burrows-Wheeler-Transform}
	\acro{BFGS}{Broyden-Fletcher-Goldfab-Shanno}
	\acro{KKT}{Karush-Kuhn-Tucker}
	\acro{WFC}{Weighted Frequency Counting}
	\acro{MTF}{Move-to-Front}
	\acro{LP}{Laplace}
	\acro{SAKDC}{Swiss Army Knife Data Compression}
	\acro{SQP}{Sequential Quadratic Programming}
\end{acronym}

\title{Combining non-stationary prediction, optimization and mixing for data compression}

\author{
	\IEEEauthorblockN{Christopher Mattern}
	\IEEEauthorblockA{
		Fakultät für Informatik und Automatisierung\\
		Technische Universität Ilmenau\\
		Ilmenau, Germany \\
		christopher.mattern@tu-ilmenau.de
	}
}


%


\maketitle


\begin{abstract}
In this paper an approach to modelling non-stationary binary sequences, i.e., predicting the probability of upcoming symbols, is presented. After studying the prediction model we evaluate its performance in two non-artificial test cases. First the model is compared to the \acl{LP} and \acl{KT} estimators. Secondly a statistical ensemble model for compressing \acl{BWT} output is worked out and evaluated. A systematic approach to the parameter optimization of an individual model and the ensemble model is stated. 

\end{abstract}


%
\IEEEpeerreviewmaketitle


\section{Introduction} \label{sec:intro}

\subsection{Background}

Sequential bitwise processing plays a key role in several general-purpose lossless data compression algorithms, including \ac{DMC} \cite{dmc}, \ac{CTW} \cite{ctw_95} and the recently emerging \ac{PAQ} \cite{cm_paq6,hbdc} family of compression algorithms. All of these algorithms belong to the class of statistical data compression algorithms, which split the compression phase into modelling and coding. A statistical model assigns probabilities to upcoming symbols and these are translated into corresponding codes. Assigning a high probability to the actually upcoming symbol leads to a short encoding, thus producing compression. The ideal code length corresponding to a prediction can closely be approximated via \ac{AC} \cite{mtc}. Hence improving prediction accuracy is crucial for compression.

Recently, \ac{PAQ}-based compression algorithms have been of high public interest, due to the enormous compression achieved. Unfortunately, there is little up-to-date literature on the internals of the involved algorithms \cite{cm_paq6, hbdc, cm_cmidc}. \ac{PAQ} compression algorithms combine multiple binary predictors and are characterized by low processing speed and the best compression rates in multiple benchmarks up to date. Ensemble prediction has previously been applied successfully in other areas of research, e.g., time series forecast and classification \cite{ens_ex1,ens_ex2,ens_ex3}, and form a promising direction of research. In the field of compression an ensemble approach is often called \ac{CM}.

The most elementary task of the prediction model is sequential probability assignment, i.e., predicting the probability distribution $P(y_{k+1} | y_{k} y_{k-1} \dots y_1)$ of the upcoming symbol $y_{k+1}$ based on the already encountered sequence $y_1 y_2 \dots y_k$ over a finite alphabet $\Sigma$. Such a task typically arises when working with context models. A finite number of symbols preceding $y_{k+1}$ can be used to condition the probability, which leads to finite context modelling \cite{mtc}. The finite context, e.g.,  the character immediately preceding the current one, splits the source sequence into sub-sequences. Th\underline{e}s\underline{e} ar\underline{e} oft\underline{e}n call\underline{e}d cont\underline{e}xt histori\underline{e}s. For instance the context history of the context ``e'' (underlined) regarding the last sentence is ``s\_\_ndxs'', an underscore represents a space symbol. This work focuses on binary alphabets, $\Sigma = \{0, 1\}$ and uses the convention $p_k = P(y_k=1)$. 
\subsection{Previous work}
Experiments have shown that a local adaption of the computed statistics during modelling typically improves compression. Thus more recent observations are of higher importance for probability assignment \cite{mtc,ac_piac}.
This observation was made more or less accidentally due to limited calculation precision, which lead to a periodic rescaling of character counts \cite{mtc}. Previous work investigated the effect of scaling \cite{ac_aac} and pointed out an approximate probability estimation model for binary sequences based on exponential smoothing \cite{ac_piac}. Another aspect is the presence of noise within observations.
An imperfect choice of conditioning contexts will lead to observations within context histories, which deviate from the governing probability distribution. We consider such events as outliers or simply noise.
A recent work \cite{seqpr} studied the effect of a limited probability interval, i.e., $\theta \in [\alpha, \beta] \subset [0, 1]$ along with the estimation of the parameter $p_k = \theta = \text{const}$ regarding a series of \ac{iid} binary random variables.
A limited probability interval can be explained by viewing an observed sequence as the outcome of the transmission of the ``true'' sequence through a noisy channel (i.e., an extension to the original source model). Results indicate that having knowledge about the parameters $\alpha$ and $\beta$ can lead to significant improvements in compression for short to medium sized sequences.
Thus using the restriction $p_k \in [\alpha, \beta]$ can represent a countermeasure for noisy observations.

\subsection{Our contribution}

The previous section explained the aspects of observation recency and observation uncertainty. Based on these ideas we enhance a standard approach for sequential binary prediction and introduce a new prediction model. 
We further employ our prediction model to construct a new ensemble compression algorithm. This compression algorithm is intended to be used as a second step algorithm in \ac{BWT} based compression. Both, the sequential prediction model and the ensemble compression algorithm, contain constants (fixed during compression or decompression), which influence the probability estimation and the compression.
We denote such constants as parameters of the algorithm or parameters of the prediction model (which should not be confused with parameters of a distribution).
In the general setting there are no simple rules for choosing the (unknown) parameters. Among the set of feasible parameters, we want to chose the parameters according to a certain objective. In data compression this objective is the minimization of the size of the compressed output.
Most of the parameter optimization in the area of data compression was carried out using ad-hoc hand-tuning, e.g.,  \cite[p. 4]{bwt_wm01}, \cite[p. 6]{ppm_vlc} and \cite[p. 4]{volf_sw}. In this work we want to introduce systematic approaches to automated parameter optimization, since these will improve the compression performance compared to ad-hoc hand-tuning.

We distinguish two versions of automatic parameter optimization in compression, which we call offline and online optimization.
Given a training data set the models' parameters can be fitted once and remain static during future usage (offline optimization). This approach requires a carefully chosen set of training data. Since the optimization takes place only once and not prior to every compression pass there are no significant restrictions on the amount of data and the associated processing time.
On the other hand, adding an initial optimization pass prior to compression and saving the parameters along with the compressed data refers to an online approach. However, there are more severe restrictions on the utilized resources.
We consider a situation in which the optimization pass requires orders of magnitude more time than the actual (de-)compression process impractical for online optimization.
In this work we focus on online optimization and incorporate an automated optimization pass into the ensemble model mentioned above. Coupling online optimization and statistical compression leads to asymmetric statistical compression, a new family of statistical compression algorithms. Without optimization such algorithms are typically symmetric, since modelling and coding is required during compression and decompression. Similar approaches to asymmetric algorithms exist in the field of audio compression \cite{optimfrog}.


There is another non-obvious benefit in using optimization. Assume an algorithm $A$ achieves a certain compression rate using an ad-hoc parametrization. A computationally cheaper algorithm $B$ produces compression comparable to $A$ along with optimized parameters. Thus the compression time is reduced when $A$ is replaced by $B$. This argument holds especially for offline optimization: The time required for optimization does not need to be included in the compression time, since optimization is only carried out once.

The remaining part of this work is divided into four further sections. First we present a new elementary, binary prediction model, its application to non-binary alphabets and an approach to ensemble prediction. Section \ref{sec:optimization} briefly summarizes iterative numeric optimization and its application to the presented modelling algorithms. Afterwards Section \ref{sec:experiments} evaluates the model components' performance and the impact of optimization.


\section{Modelling} \label{sec:modelling}

\subsection{Elementary prediction} \label{sec:el_model}

As previously mentioned in Section \ref{sec:intro} the most essential task is to estimate the probability distribution \mbox{$p_{n+1} = P(Y_{n+1}=1 | B_n = b_n)$} given the series of binary random variables $B_n = Y_1 Y_2 \dots Y_n$ and an instance $b_n = y_1 y_2 \dots y_n \in \{0, 1\}^n$, where $m$ out of $n$ bits are one. 
Assuming \ac{iid} random variables $Y_k$,
i.e., $p_k = \theta$ for all $k$ and some fixed $\theta \in [0, 1]$, one can calculate the probability of a given outcome $b_n$ via
\begin{equation}
	P(B_n = b_n | \theta ) = \prod_{k=1}^n P(Y_k=y_k) = \theta^{m} (1-\theta)^{n-m} .
\end{equation}
When $b_n$ is fixed an estimation $\hat \theta$ of $\theta$ can be obtained via maximizing $P(\theta | b_n)$, or via minimizing the entropy
\begin{eqnarray}
	H(\theta | b_n) &=& -\sum_{k=1}^n \log P(Y_k=y_k) \label{eq:seq_entropy1} \\
	       &=& -m \log \theta + (n-m) \log(1-\theta). \nonumber
\end{eqnarray}
Note that logarithms are to the base two. The result of minimizing \eqref{eq:seq_entropy1} is the well-known maximum likelihood estimator $\hat \theta = m/n$.
Equation \eqref{eq:seq_entropy1} is rewritten to yield
\begin{equation}
	H(b_n) = - \sum_{k=1}^n \left( y_k \log \theta + (1-y_k) \log (1-\theta) \right) \label{eq:seq_entropy}.
\end{equation}
Since we assume that the coding cost of more recent events is of higher importance, 
we modify \eqref{eq:seq_entropy} to become a weighted entropy (cf. \cite{ac_aac})
\begin{equation}
	H_w(b_n) = -\sum_{k=1}^n c_k \left( y_k \log \theta + (1-y_k) \log (1-\theta) \right) \label{eq:wt_seq_entropy},
\end{equation}
where $0 < c_1 < c_2 < \dots < c_n$ is some weight sequence.
In this way, the value of $\theta$ is strongly linked to more recent observations (steps $n, n-1, \dots$).

Next we address the aspect of observation uncertainty, similar to \cite{seqpr}. The observations $y_k$ are viewed to be the outcome of a binary symmetric channel.
On the transmitter side the outcome $y_k$ of a binary random variable $Y_k$ is sent through the channel.
The receiver observes a corrupted bit $1-y_k$ with a probability $\varepsilon$, i.e., the outcome of a binary random variable $X_k$. Summarizing
\begin{eqnarray}
	P(X_k = y_k | Y_k = y_k) &=& 1-\varepsilon \label{eq:bsc},\\
	P(X_k \neq y_k | Y_k = y_k) &=& \varepsilon \nonumber
\end{eqnarray}
holds for some $0 \leq \varepsilon \leq 0.5$. Thus \eqref{eq:wt_seq_entropy} is modified to become the expected, weighted entropy
\begin{eqnarray}
	\overline H_w(b_n) &=& - \sum_{k=1}^n c_k \left( \delta_k \log \theta + (1-\delta_k) \log(1-\theta) \right), \\
	\delta_k &=& (1-\varepsilon) y_k + \varepsilon (1-y_k) \nonumber,
\end{eqnarray}
since we can only observe the receiver side.
We assume that the statistical properties of the bit sequence do not change rapidly (i.e., $p_{n+1} \approx p_n$) and approximate $p_n$ using the solution of the minimum-entropy problem
\begin{equation}
	p_{n+1} \approx p_n = \arg \min_\theta \overline H_w(b_n),
\end{equation}
which results in 
\begin{eqnarray}
	p_{n+1} &\approx& \frac{\sum_{k=1}^n c_k \delta_k}{\sum_{k=1}^n c_k} \label{eq:pr_est1} \\
		&=& \varepsilon + (1-2\varepsilon) \frac{\sum_{k=1}^n c_k y_k}{\sum_{k=1}^n c_k}. \nonumber 
\end{eqnarray}
Thus modelling uncertainty via \eqref{eq:bsc} restricts the probability interval to be $[\varepsilon, 1-\varepsilon]$. As a side effect the problem of assigning a probability to the opposite bit $y_{k+1} = 1-b$ when processing a deterministic sequence $y_1 = y_2 = \dots = y_k = b \in \{0, 1\}$ is solved. The source model discussed above contains several (generally unknown) parameters - the weight sequence and $\varepsilon$. In order to use the source model for prediction these parameters have to be chosen. A bad choice leads to redundancy during coding, e.g., in some step $k$ the actual value of $p_k$ could be located outside of the restricted probability interval, but the model is only able to assign values in $[\hat \varepsilon, 1-\hat \varepsilon]$ depending on the estimated parameter $\hat \varepsilon$.

\subsection{Efficient approximations} \label{sec:el_model_approx}

Equation \eqref{eq:pr_est1} can already be utilized to obtain a probability estimation given a weight sequence and $\varepsilon$. However, from a practical point of view and as a matter of convenience an estimation should be calculated incrementally, hence we select an exponentially decaying weight sequence
\begin{equation}
	c_k = \lambda^{n-k}, ~1 \leq k \leq n ,
\end{equation}
with $\lambda \in (0, 1]$. Equation \eqref{eq:pr_est1} becomes
\begin{equation}
	p_{n+1} = \frac{S_{n+1}}{T_{n+1}},
\end{equation}
where
\begin{eqnarray}
	S_{n+1} &=& \lambda S_{n} + \delta_n, \\
	T_{n+1} &=& \lambda T_{n} + 1,
\end{eqnarray}
which can be reformulated to yield an adjustment proportional to the prediction error
\begin{eqnarray}
	p_{n+1} &=& p_n + \frac{1}{T_{n+1}}( \delta_n - p_n ) \label{eq:pr_est2}. 
\end{eqnarray}
Initially we have $p_0 = 0.5$ and $T_0=0$. Note that the sequence $T_n$ is a geometric series and therefore
\begin{equation}
	 T_n \xrightarrow[n\rightarrow\infty]{} \frac{1}{1-\lambda}. \label{eq:Tn_series}
\end{equation}
For a very long sequence exponential smoothing can be used as an approximation of \eqref{eq:pr_est2}, i.e,
\begin{eqnarray}
	p_{n+1} &=& p_n + (1-\lambda) (\delta_n - p_n), \label{eq:pr_est_approx} \\
	&=& \lambda p_n + (1-\lambda) \delta_n. \nonumber
\end{eqnarray}
Depending on the computational resources different approximations seem acceptable:
\begin{itemize}
	\item \textbf{Exact model $\mathbf{M_1}$.} An estimator state is $(p_n, T_n)$, computed according to \eqref{eq:pr_est2}.
	\item \textbf{Exponential smoothing $\mathbf{M_2}$.} The state is given by $(p_n)$ and is updated following \eqref{eq:pr_est_approx}. Selecting $1-\lambda = 2^{-l}$, \mbox{$l \in \mathbb{N}$} results in a very efficient calculation using bit shifts and additions/subtractions only.
\end{itemize}
Note that $M_1$ can be approximated more closely by imposing an upper limit on $T_n$, or $n$, respectively. This yields a state $(p_n, n')$, with $n'=\min(n,\overline n)$ for a threshold $\overline n$. The values of $1/T_n$ are found using a lookup table and $T_{\overline n} \approx T_\infty$ is set according to \eqref{eq:Tn_series}. All approximations described above share the same parameters $\lambda$ and $\epsilon$.

\subsection{Alphabet decomposition and context modelling} \label{sec:deco}

\begin{figure}[!t]
	\hrule
	\smallskip
	\centering
	\small{
	\begin{tabular}{cccccc}
	 & $s_1$ & $s_2$ & $s_3$ & $s_4$ & \\
	Bit & 72 (H) & 101 (e) & 108 (l) & 108 (l) &  \dots \vspace{.2cm} \\ 
		$y_8$ & 01001000 & \underline{01100101} & \textbf{0}1101100 & 01101100 & \dots \\ 
		$y_7$ & 01001000 & \underline{01100101} & \underline{0}\textbf{1}101100 & 01101100 & \dots \\
		\dots\\
		$y_1$ & 01001000 & \underline{01100101} & \underline{0110110}\textbf{0} & 01101100 & \dots \vspace{.2cm} \\
		$y_8$ & 01001000 & 01100101 & \underline{01101100} & \textbf{0}1101100 & \dots \\
		$y_7$ & 01001000 & 01100101 & \underline{01101100} & \underline{0}\textbf{1}101100 & \dots \\
		\dots 			
	\end{tabular}}
	\smallskip
	\hrule
	\caption{The decomposed symbol $s_3$ is encoded in eight consecutive binary steps (current bit is boldface) using an order-$1$ context (underlined), i.e., the predictions $P(y_8=1\mid s_2=101)$, $P(y_7=1\mid y_8=1, s_2=101)$, \dots, $P(y_1=1\mid y_2y_3\dots y_8=0110110, s_2=101)$ need to be calculated. After encoding $s_3$, $s_4$ can be processed in the same fashion. }
	\label{fig:deco}
\end{figure}

The previous section dealt with the modelling of a binary alphabet. In general the compression algorithms work on $n$-ary alphabets $\Sigma$, typically $|\Sigma|=2^8$. Hence bitwise processing requires an alphabet decomposition, i.e., a mapping $\mathrm{code}: \Sigma \rightarrow \{0, 1\}^{+}$ and \mbox{$\mathrm{len}: \Sigma \rightarrow \mathbb{N}$} to indicate the code length. Without loss of generality we may assume that $\Sigma = \{ 0, 1, 2, \dots, m-1 \}$.
Within this work we use a fixed decomposition, which we call ``flat decomposition'', i.e., $\mathrm{code}(s) = \mathrm{bin}(s)$ (e.g., $\mathrm{code(65)} = 01000001$) and $\mathrm{len}(s) = L = 8$ for every symbol $s\in\Sigma$. Modelling the probability distribution of $s$ is split into $\mathrm{len}(s)=L$ consecutive steps
\begin{equation}
	P(s) = P(y_L) P(y_{L-1}|y_L) \dots P(y_1 | y_2 y_3 \dots y_L).
\end{equation}
Working with conditional probabilities increases the prediction accuracy. A natural choice are order-$N$ contexts, which have successfully been applied to text compression \cite{hbdc, mtc}. An order-$N$ context consists of the last $N$ characters immediately preceding the current one. 
Figure \ref{fig:deco} illustrates the bitwise modelling process using an order-1 context.
Depending on the underlying data other choices can be reasonable as well, e.g., the neighbouring pixels in image compression \cite{ppm_exe}, \cite{ppm_img}.

\subsection{An ensemble predictor} \label{sec:ens_model}

Section \ref{sec:intro} mentioned the successful application of ensemble models in other areas. In the area of compression such techniques are known \cite{mtc}, but there has been less interest in directly applying them. Such techniques allow multiple models to contribute with their advantages without cumulating their disadvantages \cite{cm_cmidc}. 
During modelling a probability must be calculated for each alphabet symbol $s\in\Sigma$, hence combining $M$ models roughly requires $M \cdot |\Sigma|$ operations. On the other hand bitwise processing just requires $M \cdot \overline L$ operations on average, where $\overline L$ is the average code length. Without making further assumptions about symbol frequencies, i.e., applying the decomposition described in Section \ref{sec:deco}, we get $\overline L=L = \lceil \log |\Sigma| \rceil$. 
An advantage of \ac{CM} compared to \acf{PPM} is that it does not need to handle symbols, which did not appear in the current context, in a special way \cite{skibinski}. \ac{PPM} indicates the presence of such a situation using an artificial escape symbol, whose probability needs to be modelled in every context. However, such situations may add redundancy, since there is code space allocated for possibly never appearing symbols. This issue can be crucial for \ac{PPM} \cite{ppm_ii}. A disadvantage of \ac{CM} is the requirement of multiple models simultaneously, which has heavy impact on processing speed and memory requirements.

We now describe the outline of our approach to ensemble prediction, or \ac{CM} respectively. It is based on a source switching model \cite{cm_cmidc}. Consider a set of $M$ sources and a probabilistic switching mechanism, which selects source $i$ with a probability of $w^i_k$ (in step $k$) where $\sum_{i=1}^M w^i_k=1$. Note that the switching model should not be confused with Volf's switching method \cite{volf_sw}, which is based on switching between \emph{source coding algorithms} rather than constructing an ensemble source model. In its current state $x^i_k$ the selected source emits a one-bit with the probability \mbox{$p_k^i = P(Y_k=1|x^i_k)$}. Afterwards a state transition takes place for each source resulting in the next state $x^i_{k+1}$. In an analogous fashion the switching probabilities may vary, i.e., these may depend on a state, too. Summarizing the probability of a one-bit in step $k$ is
\begin{equation}
	p_k = \sum_{i=1}^M w^i_k p^i_k. \label{eq:lin_mix}
\end{equation}
Thus the assumption (or approximation) of a switching source results in a linear ensemble prediction (linear mixing).
Unfortunately, normally no information about the internals of the source (e.g., involved states and transitions) or the characteristics of the probability assignment is available. The assignment is up to the designer.

\subsection{Applications and test cases} \label{sec:model_app}

\begin{figure}[!t]
	\hrule
	\smallskip
	\centering
	\small \texttt{eiehdnkleeeeeeeeeeeiiiiiiiiiiiiyyeeeeei\\
		iieeeeiieeeeiieeeeeeeeeeeeeeyiyyyyiiiii\\
		iiyyyiyyiyyiiyyyyiyyyyyiyyyeeeeeeeeeeee\\
		eeeeeeeeeeeeeeeeeyyeeeeeeceeeeeeeeeieee\\
		eeeeehhohhhhheeeeeeeeeeeeeeeeeeieeeeeee\\
		eeeeeeeeeeeeeeeeeeyeeeeeeeeeeeeeeeeeeee\\
		iiiieeeeeeeeeeeeeeeeeeeeeeeeeeehheeeeee\\
		eeeeeeeeeeeeeeeeeieeeeeeeeeeeeeeeeeeyee\\
		eeeeeeeeeeeyeeeeeeeeeeeeeeeeeeeeeeeeeee\\
		eeeeeeeeeeeeeeeeeeeeeeeeeeeeeeeeeeeeeee\\		
	}
	\smallskip
	\hrule
	\caption{A typical example of \ac{BWT} output taken from \textit{book1} (Calgary Corpus). }
	\label{fig:bwt_output}
\end{figure}

\subsubsection*{Single model}
To examine the prediction model described in Sections \ref{sec:el_model} and \ref{sec:el_model_approx} we will compare its performance to the well-known \acf{LP}- and \acf{KT}-estimators \cite{ctw_95, eoit} with scaling \cite{ac_aac}.

\subsubsection*{Ensemble model} 
For testing the ensemble approach we introduce a simple ensemble compression algorithm intended as a second step algorithm in \ac{BWT} based compression.
\ac{BWT} sorts the characters in its input by context, hence it groups similar contexts together \cite{bwt_94}. Since these contexts are often succeeded by the same characters \ac{BWT} output mostly consists of long interleaved runs of characters, see Fig. \ref{fig:bwt_output}. Such sequences can be modelled as non-stationary \cite{bwt_wm01}. We model the \ac{BWT} output as the outcome of a switching source, which consists of two individual non-stationary sources. One source randomly emits characters independent of the previous sequence (order-0), this is intended to model interruptions in a single characters run. A second source emits characters based on the character immediately preceding the current position (order-1).
In contrast to the first source it is intended to model the long runs of identical characters.
The individual models are implemented using the binary predictors described in Section \ref{sec:modelling}. We assume the switching probabilities to be constant, i.e., $w^2_k=1-w^1_k = \omega \in [0, 1]$.

Each individual model presented in Section \ref{sec:el_model_approx} has two parameters $\lambda$ and $\varepsilon$. The previously described \ac{BWT} postprocessor has five parameters, $\lambda_1$, $\varepsilon_1$, $\lambda_2$, $\varepsilon_2$ and $\omega$, respectively. Following these observations the next section will provide a way of optimizing the parameters.


\section{Optimization} \label{sec:optimization}

\subsection{Iterative numeric optimization}

\begin{figure}[!t]
	\hrule
	\smallskip
	\centering
	\small
	\begin{algorithmic}
		\STATE $\bold x_0 \leftarrow \text{initial estimation}$
		\STATE $k \leftarrow 0$
		\REPEAT
			\STATE compute a search direction $\bold d_k$ along which $f$ decreases
			\STATE perform a line search $\alpha_k \gets \arg \min_\alpha f(\bold x_k + \alpha \bold d_k)$
			\STATE update the solution $\bold x_{k+1}\gets \bold x_k + \alpha_k \bold d_k$
			\STATE next step $k \gets k+1$
		\UNTIL{stopping condition met}
	\end{algorithmic}
	\hrule
	\caption{Basic outline of an iterative numeric minimization algorithm.}
	\label{fig:num_opt}
\end{figure}

We decompose a model into its structure and parameters.
Improving the model structure is a task which is typically carried out by humans. Model parameters can be fitted automatically to a typical training data set. There are different approaches, depending on the optimization target. A differentiable optimization target allows the usage of local search procedures, for instance Newton's Method, see standard literature on these well-known techniques, e.g.,  \cite{bertsekas}. When no derivative information is available (i.e., a non-differentiable optimization target) or the search space is highly multimodal other stochastic search techniques should be preferred, see e.g., \cite{schaefer}. In our setting we want to minimize the average code length $f$, depending on the \mbox{parameters $\bold x$} of the prediction model
\begin{equation}
	\min_{\bold{ \underline x} \leq \bold x  \leq \bold{ \overline x} } f(\bold x),
\end{equation}
where $f(\bold x)$ is given by a modification of \eqref{eq:seq_entropy}
\begin{equation}
	f(\bold x) = -\frac {1}{n} \sum_{k=1}^n \left(y_k \log p_k(\bold x) + (1-y_k) \log(1-p_k(\bold x)) \right). \label{eq:opt_tgt}
\end{equation}
Here boldface symbols indicate matrices or vectors. The parameter search should take place within the hypercube formed by the inequality constraints $\bold x \in [\bold {\underline x}, ~\bold {\overline x}] \subset \mathbb{R}^N$. In this work we want to focus on derivative-based optimization techniques based on Quadratic Programming, since $f$ is differentiable. Figure \ref{fig:num_opt} shows the typical outline of such an optimization procedure.
The models described in the previous Section span a low-dimensional search space, e.g., $\bold x = (\lambda_1, \varepsilon_1, \lambda_2, \varepsilon_2, \omega)^T \in \mathbb{R}^5$. Opposed to the small number of parameters a function evaluation is, depending on the amount of training data, time consuming. It requires to run the corresponding model along with the calculation of derivatives. Since we want to use an online-optimization approach, the ``training data'' is the data to be actually compressed, i.e., we know it prior to optimization.

\subsection{Estimating the search direction}

Consider a quadratic approximation $f(\bold x_k + \bold d_k)$ of the target function $f$ as a result of the Taylor-expansion at $\bold x_k$
\begin{equation}
	f(\bold x_k + \bold d_k) \approx f(\bold x_k) + \bold \nabla f(\bold x_k)^T \bold d_k + \frac{1}{2} \bold d_k^T \bold \nabla^2 f(\bold x_k) \bold d_k. \label{eq:opt_taylor}
\end{equation}
Differentiating \eqref{eq:opt_taylor} in $\bold d_k$ and solving for its roots yields a search direction
\begin{equation}
	\bold d_k = \underbrace{-\bold \nabla^2 f(\bold x_k)^{-1}}_{\bold S_k} \bold \nabla f(\bold x_k). \label{eq:opt_dir}
\end{equation}
The matrix $\bold S_k$ can either be estimated iteratively or computed directly.
Given a valid point $\bold x_k \in [\bold {\underline x}, ~\bold {\overline x}]$ a step towards $\bold d_k$ might lead to a violation of the constraints. Hence the constraints influence the computation of $\bold d_k$. In order to calculate a feasible direction we adopt a slight modification of the method in  \cite{boxopt}, which we will now summarize briefly.

First the index set of binding constraints 
\begin{equation}
	I_k = \underbrace{I'_k(-\nabla f(\bold x_k))}_{I^1_k} ~\cup~ \underbrace{I'_k( -\bold S'_k \bold \nabla f(\bold x_k) )}_{I^2_k}
\end{equation}
is identified depending on
\begin{equation}
	I'_k(\mathbf \delta) = \left \lbrace i ~|~ x^i_k = \underline x^i_k \wedge \delta_i < 0 \vee  x^i_k = \overline x^i_k \wedge \delta_i > 0 \right \rbrace.
\end{equation}
An element $s^{ij'}_k$ of $\bold S'_k$ is given by
\begin{equation}
	s^{ij'}_k = \begin{cases}
		s^{ij}_k &, i,j \notin I^1_k \\
		0 &, \text{otherwise}
	\end{cases}
\end{equation}
depending on the elements $s^{ij}_k$ of $\bold S_k$. With $\delta_i$ we denote the $i$-th component of $\mathbf \delta$, the same holds for $\underline x_i$ and $\overline x_i$, respectively. The set $I_k(\mathbf \delta)$ contains the indices of blocked directions, i.e., $x_i$ is located on a constraint boundary and $\delta_i$ points towards the constraint. Constraints contained in $I^1_k$ block movements along the directions fulfilling the \ac{KKT} conditions and $I^2_k$ blocks movements, which would leave the feasible region due to the linear transform described by $\bold S_k$. Finally given $I_k$ the search direction is obtained via
\begin{equation}
	\bold d_k = - \bold S''_k \bold \nabla f(\bold x_k)
\end{equation}
and
\begin{equation}
	s^{ij''}_k = \begin{cases}
	s^{ij}_k &, i,j \notin I_k \\
	0 &, \text{otherwise}
	\end{cases} .
\end{equation} 

During the optimization of the parameters of a single model, we compute the gradient $\bold \nabla f(\bold x)$ and $\bold S_k = -\nabla^2 f(\bold x)^{-1}$ directly. When carrying out the experiments for the ensemble model this turned out to be too expensive computationally to be practical for our purposes.
Instead of computing $\bold S_k$ we used the \ac{BFGS} approximation in conjunction with the Sherman-Morrison formula \cite{bertsekas, boxopt} resulting in a Quasi-Newton step.

\subsection{Line search}

According to Fig. \ref{fig:num_opt} an estimation of the step length is the next step in the optimization procedure. A step along $\bold d_k$ can still leave the feasible region, when stepping too far. There is an upper limit $\overline \alpha_k$ of $\alpha$ imposed by the non-binding constraints
\begin{eqnarray}
	\overline \alpha_k &=& \min \left(\lbrace 1 \big\rbrace \cup \lbrace \tfrac{z^i_k - x^i_k}{d^i_k} ~\vline~ i \notin I_k  \rbrace \right),\\
		z^i_k &=& \begin{cases}
			\underline x_i &, d^i_k < 0 \\
			\overline x_i &, d^i_k > 0 
		\end{cases} .
\end{eqnarray}
In the case of an approximation of $\bold S_k$ the line search was carried out using quadratic interpolation. The derivative information of
\begin{equation}
	\phi_k(\alpha) = f(\mathbf x_k + \alpha \mathbf d_k)
\end{equation}
is already available at $\alpha=0$. Due to the calculation of $\bold \nabla f(\bold x_k)$ and $\bold d_k$ we get
\begin{equation}
	\phi_k'(0) = \bold d_k^T \nabla f(\bold x_k).
\end{equation}
Now a value $\beta \in (0, \overline \alpha_k]$ fulfilling $\phi_k(\beta) \geq \phi_k(0)$ is located. 
The minimum of the interpolation polynomial is given by
\begin{equation}
	\gamma = \frac{1}{2} \frac{\phi_k'(0) \beta^2}{\phi_k'(0) \beta - (\phi_k(\beta)-\phi_k(0))}.
\end{equation}
If $\phi_k$ is decreased sufficiently, i.e.,
\begin{equation}
	\phi_k(\gamma) \leq \phi_k(0) + c \gamma \phi_k'(0),
\end{equation}
where $c=10^{-5}$, we set $\alpha_k = \gamma$ and the line search is finished. Otherwise $\beta$ is replaced with $\gamma$ and the process is repeated.

\subsection{Stopping condition}

The optimization algorithm stops, when all components $\nabla_i f(\bold x_k), i\notin I_k$ are in the range $[-T, T]$.
It turned out that the precision requirements are rather relaxed, $T \in [10^{-3}, 10^{-2}]$ gives satisfying results. A higher request in precision translates into compression gains typically below $0.0001$ bpc, which can be considered insignificant. The number of iterations has been limited to $50$.

\subsection{Derivatives}

To perform the optimization process it is necessary to calculate the partial derivatives, since these form the gradient and the Hessian. For reasons of convenience we 
introduce
\begin{equation}
	h(y,p) = -y \ln p - (1-y) \ln(1-p). \label{eq:opt_h}
\end{equation}
Note that here $\ln$ denotes the natural logarithm. Using this convention \eqref{eq:opt_tgt} becomes
\begin{equation}
	f(\bold x) = \frac{1}{n \ln 2}\sum_{k=1}^n h(y_k, p_k(\bold x))
\end{equation}
and a partial derivative w.r.t. $x_i$, a component of $\bold x$, is
\begin{equation}
	\frac{\partial f(\bold x)}{\partial x_i} = \frac{1}{n \ln 2}\sum_{k=1}^n \frac{\partial h(y_k, p_k(\bold x))}{\partial x_i}.
\end{equation}
Since $y \in \{0, 1\}$ we may write
\begin{equation}
	\frac{\partial^n h(y,p)}{\partial p^n} = (n-1)! \left[ -\frac{y}{p} + \frac{1-y}{1-p} \right]^n ,
\end{equation}
The first derivative
\begin{equation}
	\frac{\partial h(y,p)}{\partial x_i} = \frac{\partial h(y,p)}{\partial p}\frac{\partial p}{\partial x_i}
\end{equation}
and the second derivative
\begin{eqnarray}
	\frac{\partial^2 h(y,p)}{\partial x_i \partial x_j} = \frac{\partial h(y,p)}{\partial x_i}  \frac{\partial h(y,p)}{\partial x_j} + \frac{\partial h(y,p)}{\partial p}\frac{\partial^2  p}{\partial x_i \partial x_j} 
\end{eqnarray}
can easily be obtained.

\subsubsection*{Single model}
First the optimization of a single model is examined, i.e., $\bold x = (\lambda, \varepsilon)^T$. 
We can restate \eqref{eq:pr_est1} as
\begin{equation}
	p_k(\bold x) = \varepsilon + (1-2\varepsilon) q_k(\lambda)
\end{equation}
where
\begin{equation}
	q_{k+1}(\lambda) = q_{k} + \frac{1}{T_{k+1}}(y_{k}-q_{k}),
\end{equation}
in the case of $M_1$ or 
\begin{equation}
	q_{k+1}(\lambda) = q_{k} + (1-\lambda)(y_{k}-q_{k}),
\end{equation}
for $M_2$, cf. \eqref{eq:pr_est2} and \eqref{eq:pr_est_approx}. Thus the required partial derivatives of $p_k$ can be expressed as
\begin{align}
	\frac{\partial p_k}{\partial \lambda} &= (1-2\varepsilon) \frac{\partial q_k}{\partial \lambda}, \\
	\frac{\partial p_k}{\partial \varepsilon} &= 1-2q_k,\\
	\frac{\partial^2 p_k}{\partial \lambda^2} &= (1-2\varepsilon) \frac{\partial^2 q_k}{\partial \lambda^2}, \\
		\frac{\partial^2 p_k}{\partial \epsilon \partial \lambda} &= \frac{\partial^2 p_k}{\partial \lambda \partial \epsilon} = -2 \frac{\partial q_k}{\partial \lambda}.
\end{align}
Depending on the choice of the model, see Section \ref{sec:el_model_approx}, the term $q_k$ remains a function of $\lambda$ \eqref{eq:pr_est2}, \eqref{eq:pr_est_approx}. Utilizing the iterative nature of \eqref{eq:pr_est2} the expressions for the exact model ($M_1$) are given by
\begin{align}
	\frac{ \partial q_{k+1} }{\partial \lambda} &= \frac{\partial q_k}{\partial\lambda} - \Delta q'_k, \\
	\frac{\partial^2 q_{k+1}}{\partial\lambda^2} &= \frac{\partial^2 q_{k}}{\partial\lambda^2} + \nonumber \\ 
	&+ \frac{1}{T_{k+1}}\left[ 2 \frac{\partial T_{k+1}}{\partial \lambda} \Delta q'_k - \frac{\partial^2 T_{k+1}}{\partial\lambda^2} \Delta q_k +  \frac{\partial^2 q_{k}}{\partial\lambda^2} \right]
\end{align}
with the abbreviations
\begin{align}
	\Delta q_k &= \frac{1}{T_{k+1}} (y_k - q_k), \\
	\Delta q'_k &= \frac{1}{T_{k+1}} \frac{\partial T_{k+1}}{\partial \lambda} \Delta q_k, \\
	\frac{\partial T_{k+1}}{\partial \lambda} &= \lambda \frac{\partial T_k}{\partial \lambda} + T_k, \\
	\frac{\partial^2  T_{k+1}}{\partial \lambda^2 } &= \lambda \frac{\partial^2  T_k}{\partial \lambda^2 } + 2 \frac{\partial T_k}{\partial \lambda}.
\end{align}
Exponential smoothing ($M_2$), \eqref{eq:pr_est_approx}, yields the following expressions:
\begin{align}
	\frac{ \partial q_{k+1} }{\partial \lambda} &= \lambda \frac{ \partial q_{k} }{\partial \lambda} - (y_k-p_k), \\
	\frac{ \partial^2  q_{k+1} }{\partial \lambda^2 } &= \lambda \frac{ \partial^2  q_{k} }{\partial \lambda^2} + 2 \frac{ \partial q_{k} }{\partial \lambda}.
\end{align}
For the initial step, $k=0$, all derivatives have been initialized to be zero.

\subsubsection*{Ensemble model}
As stated in Section \ref{sec:modelling} an ensemble model consists of an order-0 and an order-1 non-stationary model (predicting $p^1_k$ and $p^2_k$) and a switching probability, or weight $\omega$. Thus a point in parameter space is $\bold x = (\lambda_1, \varepsilon_1, \lambda_2, \varepsilon_2, \omega)^T$. The expressions for calculating the gradient worked out above just need to be modified slightly. Higher order partial derivatives are estimated using \ac{BFGS}.
 The partial derivatives of \eqref{eq:opt_h} are given by
\begin{align}
	\frac{\partial h(y,p_k)}{\partial z_i} &= w_i \frac{\partial h(y,p_k)}{\partial p_k} \frac{\partial p^i_k}{\partial z_i},
\end{align}
where $z_i \in \{\lambda_1, \varepsilon_1, \lambda_2, \varepsilon_2\}$, $p_k$ is the mixed prediction in step $k$, see \eqref{eq:lin_mix}, and
\begin{equation}
	w_i = \begin{cases} 1-\omega &, i=1 \\ \omega &, i=2 \end{cases} .
\end{equation}
Finally the remaining derivative for the ensemble model is
\begin{equation}
	\frac{\partial h(y,p_k)}{\partial \omega} = \frac{\partial h(y,p_k)}{\partial p_k} (p^2_k-p^1_k).
\end{equation}


\section{Experimental evaluation} \label{sec:experiments}

\subsection{Single model}

\begin{table}[!t]
	\caption{Compression rates (Calgary Corpus) in bpc and average context history length $L$ of different order context histories for the \acl{LP}, \acl{KT} and the developed $M_1$ and $M_2$ estimators (Section \ref{sec:el_model_approx}).}
	\centering
	\begin{tabular}{|c|c|c|c|c|c|}
		\hline
		\textbf{Order} & \textbf{L} & \textbf{LP} & \textbf{KT} & $\mathbf{M_1}$ & $\mathbf{M_2}$ \\ \hline
		\textit{0} & \textit{14793.1} & 4.749 & 4.731 & \textbf{4.717} & 4.719 \\
		\textit{1} & \textit{328.5} & 3.581 & \textbf{3.525} & 3.528 & 3.554 \\
		\textit{2} & \textit{37.4} & 3.252 & 3.115 & \textbf{3.075} & 3.222 \\
		\textit{4} & \textit{5.3} & 3.965 & 3.671 & \textbf{3.442} & 3.620 \\
		\textit{8} & \textit{2.2} & 5.651 & 5.371 & \textbf{5.013} & 5.101 \\ \hline
	\end{tabular}	
	\label{tab:bitmodels}
\end{table}

\begin{figure}[!t]
	\centering
	\includegraphics[width=7.5cm]{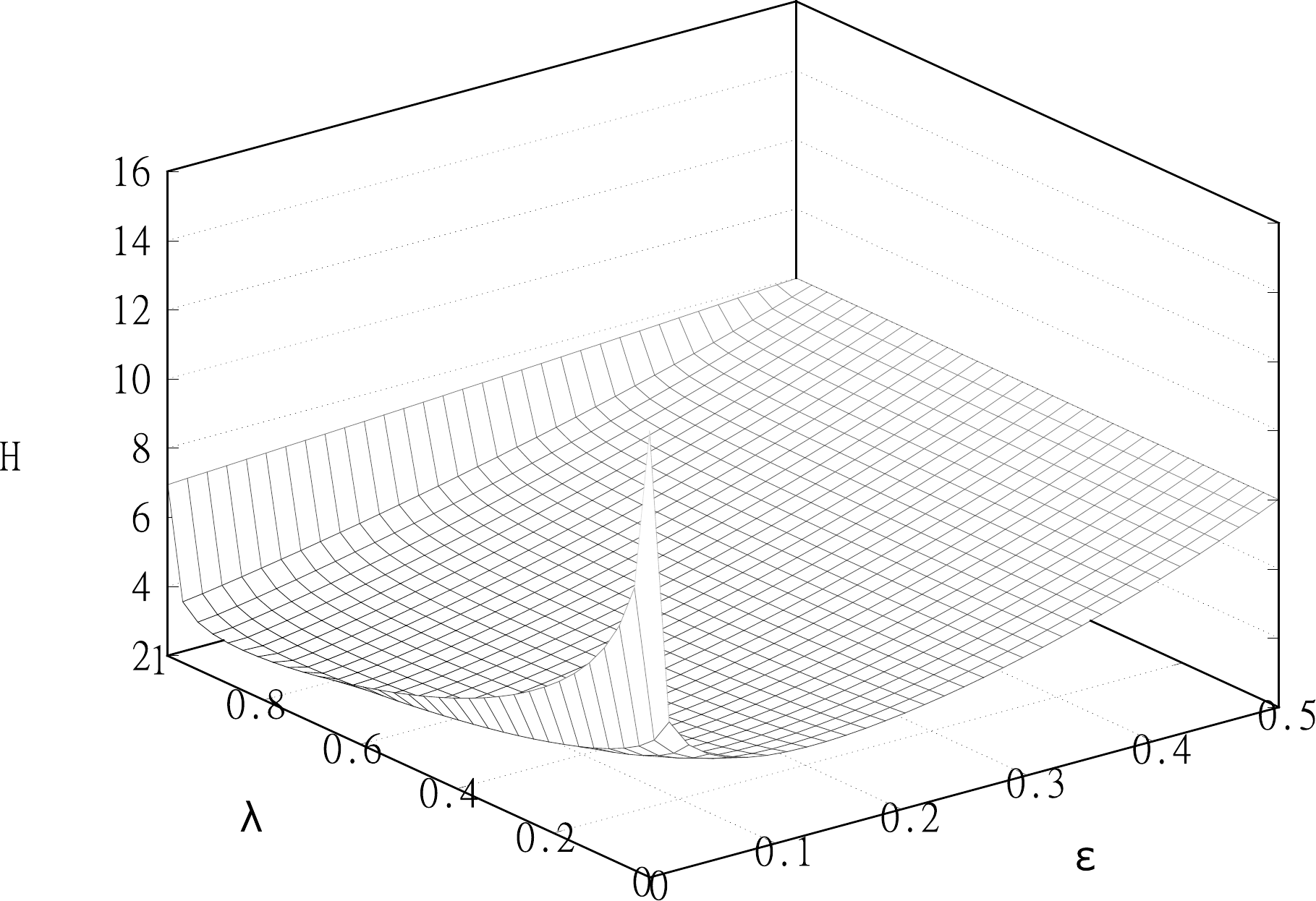}
	\caption{Entropy $H(\lambda, \varepsilon)$ of the order-2 context histories of \textit{bib}.}
	\label{fig:bitmodel1_surface}
\end{figure}

To evaluate a single prediction model we compare its compression performance against the well-known \ac{LP} and \ac{KT} estimators when forecasting conditional probabilities for different order-$N$ context models. Note that $N$ terms the number of source symbols, bytes in this case. The flat alphabet decomposition described in Section \ref{sec:deco} has been applied. The competing models are given by
\begin{equation}
	p_k = \frac{S_k + \alpha}{T_k + 2 \alpha},
\end{equation}
where $S_k$ is the frequency of a one bit, $T_k$ the total number of encountered bits in step $k$ and $\alpha$ distinguishes the \ac{LP}- ($\alpha=1$) and \ac{KT}-estimator ($\alpha=0.5$)
Whenever $T_k$ reaches a threshold $\overline T$ the frequencies $T_k$ and $S_k$ are halved (scaling). 
The parameter $\overline T \in \{1, 2, \dots, 1024 \} \cup \{ \infty \}$ was optimized for each file. Table \ref{tab:bitmodels} summarizes the average compression rates per context model for the Calgary Corpus and the average context history length. The estimator $M_1$ outperforms all other predictors, except in the case of an order-1 context, where its performance is slightly worse than a \ac{KT} estimator. The \ac{LP} estimator gives the worst overall results, probably due to the uniform prior-distribution assumed \cite{ctw_95,eoit}. Especially for short context histories \cite{seqpr}, orders 2, 4 and 8, $M_1$ improves compression compared to the competing models. Exponential smoothing, $M_2$, yields a good approximation when the context history contains at least a few hundred observations (order-0 and order-1).


Figure \ref{fig:bitmodel1_surface} depicts the typical shape of the cost function, \eqref{eq:opt_tgt}, for $M_1$. When $\lambda$ is nearly zero the influence of past observations vanishes resulting in an unstable prediction behaviour, thus bad compression. A reasonable value of $\lambda$ close to 1 gives good results. In the case of $\varepsilon \approx 0.5$ virtually no compression takes place, since the probability estimates fall within a narrow band around $0.5$. A small value of $\varepsilon$ near zero is a good choice. The estimator $M_2$ shows similar characteristics.

\begin{figure}[!t]
	\centering
	\includegraphics[width=7.5cm]{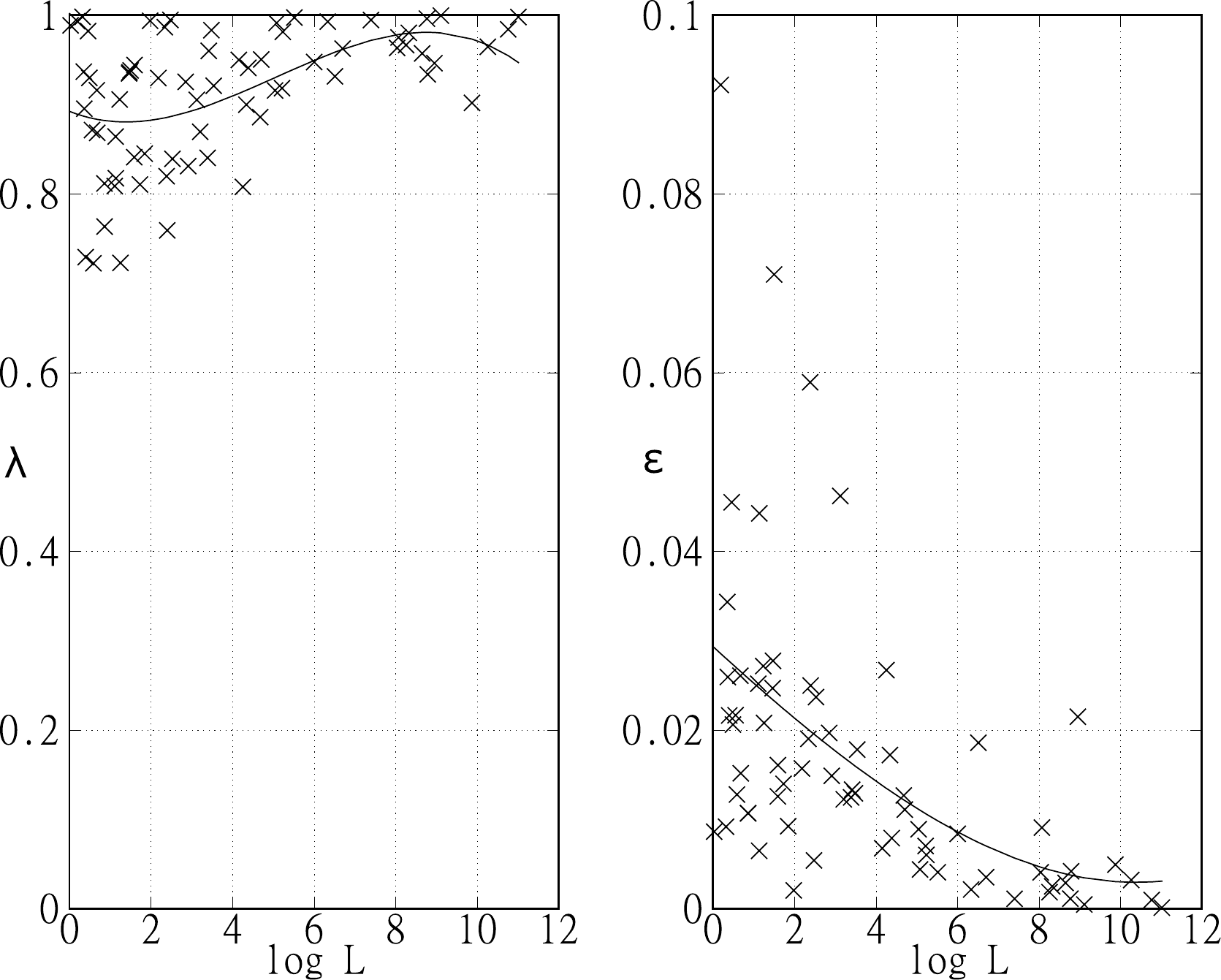}
	\caption{Parameter values $(\lambda, \varepsilon)$ and a third-order polynomial fit for $M_1$, \eqref{eq:pr_est2}, as a function of average context history length $L$.}
	\label{fig:bitmodel1}
\end{figure}

\begin{figure}[!t]
	\centering
	\includegraphics[width=7.5cm]{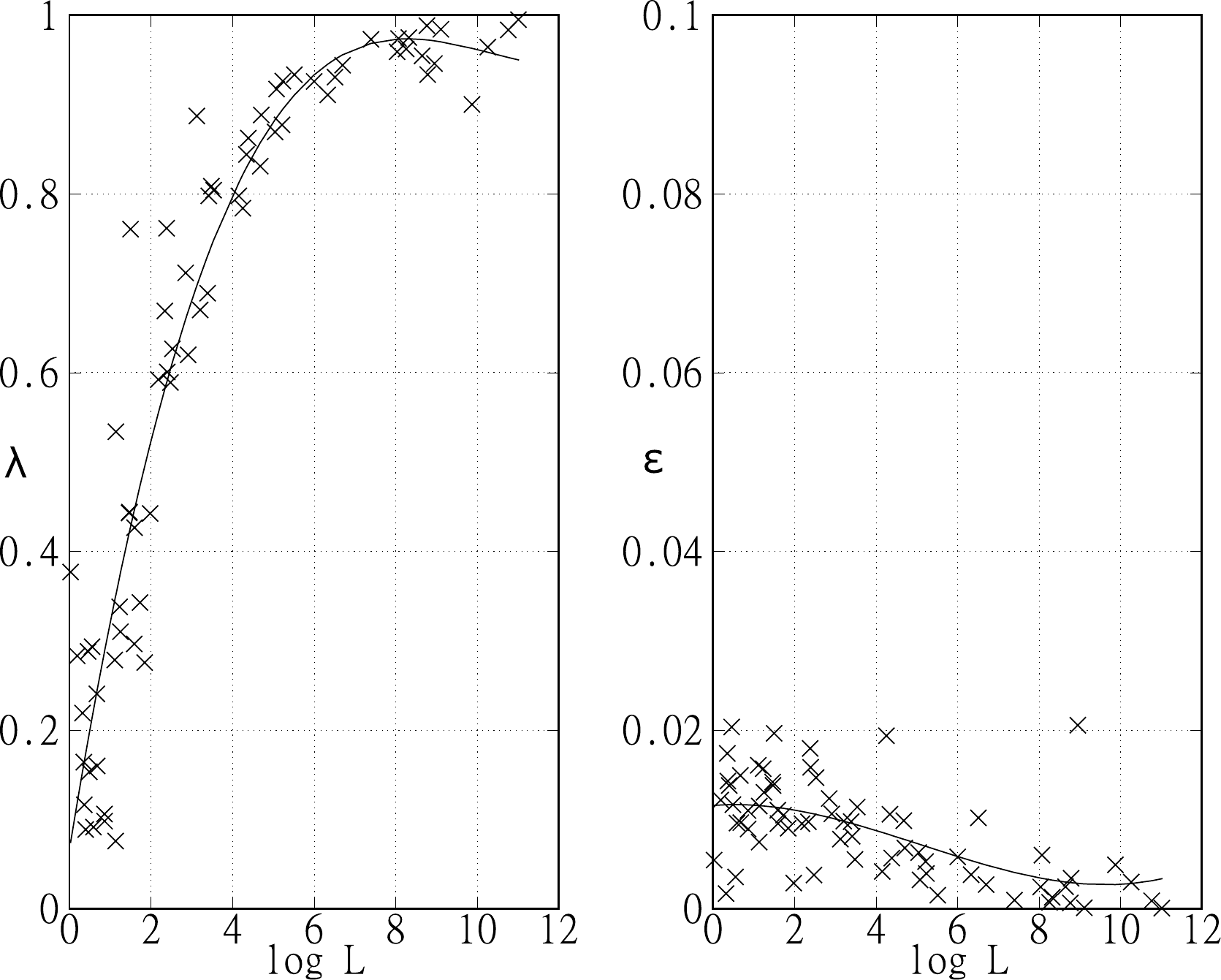}
	\caption{Parameter values $(\lambda, \varepsilon)$ and a third-order polynomial fit for $M_2$, \eqref{eq:pr_est_approx}, as a function of average context history length $L$.}
	\label{fig:bitmodel2}
\end{figure}

Figures \ref{fig:bitmodel1} and \ref{fig:bitmodel2} show the optimized values of $\lambda$ and $\varepsilon$ as a function of the context history length $L$. Shorter context histories seem to imply bigger values of $\varepsilon$ on average. This resembles the observation made in \cite{seqpr}, where it is stated that a bounded probability interval can show significant compression improvements for short sequences. The relation is more pronounced in the case of $M_1$, see Fig. \ref{fig:bitmodel1}. The parameter $\lambda$ grows as $L$ decreases, again this effect is sharper when observing $M_2$ (Fig. \ref{fig:bitmodel2}). A possible explanation is the variable adjustment proportional to the prediction error. In $M_1$ the ``adaption rate'' $1/T_{k+1}$ \eqref{eq:pr_est2} is large initially and decreases, opposed to $M_2$ where it is constant. Short sequences require a more rapid adaption, thus a constant ``adaption rate'' $1-\lambda$ \eqref{eq:pr_est_approx} should be high. We believe that the strong dependence of the parameters on $L$ in the case of very short sequence (small $L$) is triggered by the small amount of observations rather than the actual statistical properties of a context history.

\subsection{Ensemble model}

\begin{table}[!t]
	\caption{Compression rates $f_i$ (Calgary Corpus), number of cost function evaluations $\# f_i$ and gradient evaluations $\# \nabla f_i$ for $M_i$ (Sections \ref{sec:el_model_approx}) used in an ensemble model, see Section \ref{sec:ens_model}.}
	\centering
	\begin{tabular}{|l|c|c|c|c|c|c|}
		\hline
		\textbf{File} & $\mathbf{f_1} \textbf{ [bpc]}$ & $\mathbf{\# f_1}$ & $\mathbf{\# \nabla f_1}$ & $\mathbf{f_2} \textbf{ [bpc]}$ & $\mathbf{\# f_2}$ & $\mathbf{\# \nabla f_2}$ \\ \hline
		\textit{bib} & 1.945 & 16 & 10 & 1.959 & 5 & 3 \\ 
		\textit{book1} & 2.248 & 11 & 8 & 2.255 & 9 & 7 \\ 
		\textit{book2} & 1.955 & 9 & 6 & 1.962 & 4 & 3 \\ 
		\textit{geo} & 4.197 & 20 & 10 & 4.199 & 17 & 13 \\
		\textit{news} & 2.429 & 8 & 5 & 2.433 & 9 & 6 \\ 
		\textit{obj1} & 3.801 & 14 & 8 & 3.752 & 14 & 7 \\
		\textit{obj2} & 2.435 & 8 & 4 & 2.431 & 6 & 3 \\ 
		\textit{paper1} & 2.449 & 8 & 4 & 2.466 & 6 & 3 \\
		\textit{paper2} & 2.368 & 8 & 5 & 2.384 & 7 & 4 \\
		\textit{pic} & 0.712 & 35 & 19 & 0.727 & 14 & 11 \\
		\textit{progc} & 2.472 & 7 & 4 & 2.477 & 7 & 5 \\
		\textit{progl} & 1.703 & 9 & 7 & 1.721 & 8 & 7 \\
		\textit{progp} & 1.721 & 10 & 8 & 1.736 & 10 & 8 \\ 
		\textit{trans} & 1.521 & 11 & 9 & 1.528 & 11 & 9 \\ \hline
		\textit{Average} & 2.283 & 12.4 & 7.6 & 2.288 & 9.1 & 6.4 \\ \hline
	\end{tabular}
	\label{tab:cm_model}
\end{table}

In this section we simulate and compare the simple post-\ac{BWT}-stage algorithm described at the end of Section \ref{sec:modelling}. Here we want to focus on the use of optimization in an online-scenario. After the \ac{BWT}-output has been generated the model is optimized using different initial estimations
\begin{equation}
	\bold x(M) = \begin{cases}\scriptstyle{(0.67, ~0.002, ~0.91,~ 0.005, ~0.44)^T}&, M=M_1 \\ \scriptstyle{(0.72, ~0.003, ~0.96, ~0.004, ~0.44)^T}&, M=M_2\end{cases}.
\end{equation}
For decompression the optimized parameters need to be transmitted along with the compressed data. Table \ref{tab:cm_model} summarizes the results. The estimator $M_1$ shows slightly better compression than $M_2$ on average, but requires more evaluations of the cost function and the gradient for optimization. A function evaluation directly corresponds to a compression pass, a gradient evaluation is slower, since more calculations are required. Oddly, the approximation $M_2$ outperforms $M_1$ on \textit{obj1} and \textit{obj2}. This indicates that the developed model does not fit the data characteristics in this particular case. The files \textit{geo} and \textit{pic} require many more iterations than the rest of the data, the optimal values of $\bold x$ differ significantly from the rest of the corpus (and from the initial estimations), e.g., 
\begin{equation}
	\bold x_{pic}(M) = \begin{cases}\scriptstyle{(0.922, ~10^{-6},~ 0.997, ~0.002, ~0.412)^T} &, M=M_1 \\\scriptstyle{(0.931, ~10^{-6}, ~0.951, ~0.004, ~0.297)^T} &, M=M_2\end{cases}.
\end{equation}
From a practical point of view $M_2$ achieves good compression, while requiring less resources during compression and offering faster model optimization. Finally Tab. \ref{tab:bwt_coders} compares our best results to
\begin{itemize}
	\item \textbf{BW94} \cite{bwt_94} - the classical result of Burrows and Wheeler using \ac{MTF} and Huffman-Coding,
	\item \textbf{BS99} \cite{bwt_bs99} - modified \ac{MTF} and statistical modeling coupled with \ac{AC},
	\item \textbf{WM01} \cite{bwt_wm01} - parsing and encoding (via \ac{AC}) the \ac{BWT}-output and omitting second-stage transformations and
	\item \textbf{D02} \cite{bwt_wfc} - \ac{WFC}, a different post-\ac{BWT} transform, in conjunction wiht \ac{AC}.
\end{itemize}
All of the other algorithms are rather complex, since they include either special post \ac{BWT} transforms (e.g., \ac{MTF} or \ac{WFC}) and a statistical model or a sophisticated statistical model with a special parsing of the \ac{BWT} output. Our approach is very simple and straight forward, since it just consists of a simple statistical model which processes the \ac{BWT} output symbol by symbol (without a special parsing strategy). Taking the simplicity of our algorithm as a base it performs very well among the other approaches. 
The ensemble model gains 5\% over BW94 and 1.3\% over WM01. However, it compresses circa 1\% worse than BS99 and 1.5\% worse than D02. In the case of \textit{book1}, \textit{book2} and \textit{pic} the ensemble model outperforms the other algorithms, showing the benefit of an optimized non-stationary model. The main drawback of the approach is that optimization is time consuming, since the online optimization requires to compress its input between 9 ($M_2$) and 12 ($M_1$) times on average (see Tab. \ref{tab:cm_model}). But this can be neglected in a ``distribution-scenario'': compression just takes place a few times and the compressed data is distributed, e.g., over the internet and needs to be decompressed multiple times. 
\begin{table}[!t]
	\caption{Compression in bpc of various BWT-based algorithms against our best result. }
	\centering
	\begin{tabular}{|l|c|c|c|c|c|}
		\hline
		\textbf{File} & \textbf{BW94} & \textbf{BS99} & \textbf{WM01} & \textbf{D02} & \textbf{best} \\ \hline
		\textit{bib} & 2.020 & 1.910 & 1.951 & \textbf{1.896} & 1.945 \\
		\textit{book1} & 2.480 & 2.270 & 2.363 & 2.274 & \textbf{2.248} \\
		\textit{book2} & 2.100 & 1.960 & 2.013 & 1.958 & \textbf{1.955} \\
		\textit{geo} & 4.730 & 4.160 & 4.354 & \textbf{4.152} & 4.197 \\ 
		\textit{news} & 2.560 & 2.420 & 2.465 & \textbf{2.409} & 2.429 \\
		\textit{obj1} & 3.880 & 3.730 & 3.800 & \textbf{3.695} & 3.801 \\
		\textit{obj2} & 2.530 & 2.450 & 2.462 & \textbf{2.414} & 2.435 \\
		\textit{paper1} & 2.520 & 2.410 & 2.453 & \textbf{2.403} & 2.449 \\
		\textit{paper2} & 2.500 & 2.360 & 2.416 & \textbf{2.347} & 2.368 \\
		\textit{pic} & 0.790 & 0.720 & 0.768 & 0.717 & \textbf{0.712} \\ 
		\textit{progc} & 2.540 & 2.450 & 2.469 & \textbf{2.431} & 2.472 \\
		\textit{progl} & 1.750 & 1.680 & 1.678 & \textbf{1.670} & 1.703 \\
		\textit{progp} & 1.740 & 1.680 & 1.692 & \textbf{1.672} & 1.721 \\
		\textit{trans} & 1.520 & 1.460 & 1.484 & \textbf{1.452} & 1.521 \\ \hline
		\textit{Average} & 2.404 & 2.261 & 2.312 & \textbf{2.249} & 2.283 \\ \hline
	\end{tabular}
	\label{tab:bwt_coders}
\end{table}

\section{Conclusion}
In this paper a new approach to modelling non-stationary binary sequences was studied and possible low-complexity implementations have been shown. Using an iterative parameter-optimization method the parameters of the model can be fitted to training data automatically. In all test cases the new model shows a good performance compared to the \ac{LP}- and \ac{KT}-estimators. Both classic estimators are surpassed except in one case, where our models show slightly worse results. Thus in the case of compressing non-stationary data the presented models typically improves compression. Beside the usage as a binary predictor on its own an ensemble model based on two non-stationary submodels for compressing \ac{BWT} output has been designed. An alphabet decomposition is required to map the $n$-ary alphabet to a binary sequence, so the binary predictor can be used. The ensemble model contains an optimization pass prior to the actual compression. Such a simple ensemble model, together with online-optimization, shows good compression performance. Note that the ensemble model is very simple and does not apply any parsing strategies or post \ac{BWT} transforms -- it directly models symbol probabilities of plain \ac{BWT} output. In order to make such an approach more practical further steps need to be taken to speed up the optimization process. Combining multiple models in data compression is highly successful in practice, but more research in this area is needed.


\section*{Acknowledgment}

The author would like to thank Martin Aumüller, Michael Rink and Martin Dietzfelbinger for helpful suggestion and corrections, which improved the readability and made this paper easier to understand.




%
%
%

\bibliographystyle{IEEEtran}

\end{document}